\documentclass[12pt]{article}

\RequirePackage{amsmath,amsthm,amsfonts,amssymb}
\RequirePackage{graphicx}

\usepackage{float}
\usepackage{enumitem}
\usepackage{url}
\usepackage{bbm}
\usepackage{natbib}
\usepackage[nottoc,numbib]{tocbibind}
\usepackage{booktabs,caption,subcaption}
\usepackage[flushleft]{threeparttable}
\usepackage{multirow}
\newcommand{\ra}[1]{\renewcommand{\arraystretch}{#1}}
\usepackage{xr}
\usepackage{color}
\usepackage{tikz}
\usetikzlibrary{patterns}
\usepackage{comment}

\usepackage[margin=1in]{geometry}
\allowdisplaybreaks
\usepackage{setspace}
\renewcommand{\eqref}[1]{(\ref{#1})}
\newcommand{\bA}{\boldsymbol{A}}
\newcommand{\ba}{\boldsymbol{a}}
\newcommand{\bB}{\boldsymbol{B}}

\newcommand{\bD}{\boldsymbol{D}}
\newcommand{\bE}{\boldsymbol{E}}

\newcommand{\bH}{\boldsymbol{H}}

\newcommand{\bI}{\boldsymbol{I}}

\newcommand{\bj}{\boldsymbol{j}}

\newcommand{\bS}{\boldsymbol{S}}
\newcommand{\bs}{\boldsymbol{s}}

\newcommand{\bt}{\boldsymbol{t}}

\newcommand{\bV}{\boldsymbol{V}}
\newcommand{\bv}{\boldsymbol{v}}

\newcommand{\bx}{\boldsymbol{x}}

\newcommand{\by}{\boldsymbol{y}}
\newcommand{\bZ}{\boldsymbol{\Pi}}

\newcommand{\bbeta}{\boldsymbol{\beta}}

\newcommand{\bgamma}{\boldsymbol{\gamma}}

\newcommand{\bmu}{\boldsymbol{\mu}}

\newcommand{\bPsi}{\boldsymbol{\Psi}}
\newcommand{\bpsi}{\boldsymbol{\psi}}

\newcommand{\btheta}{\boldsymbol{\theta}}

\newtheorem{theorem}{Theorem}
\newtheorem{corollary}{Corollary}

\theoremstyle{definition}

\allowdisplaybreaks

\begin{document}

\begin{singlespace}

\title{\bf Fused mean structure learning in data integration with dependence}

\author{Emily C. Hector \thanks{The author thanks Dr. Andrew Whiteman for helpful discussions, Drs. Marie Davidian and Ryan Martin for reading early manuscript drafts, and Dr. Lan Luo for R code implementing the quadratic inference function sub-routine. The author is grateful to the participants of the ABIDE study, and the ABIDE study organizers and members who aggregated, preprocessed and shared the ABIDE data.}\hspace{.2cm}\\
Department of Statistics, North Carolina State University}
\date{}
\maketitle

\begin{abstract}
Motivated by image-on-scalar regression with data aggregated across multiple sites, we consider a setting in which multiple independent studies each collect multiple dependent vector outcomes, with potential mean model parameter homogeneity between studies and outcome vectors. To determine the validity of jointly analyzing these data sources, we must learn which of these data sources share mean model parameters. We propose a new model fusion approach that delivers improved flexibility, statistical performance and computational speed over existing methods. Our proposed approach specifies a quadratic inference function within each data source and fuses mean model parameter vectors in their entirety based on a new formulation of a pairwise fusion penalty. We establish theoretical properties of our estimator and propose an asymptotically equivalent weighted oracle meta-estimator that is more computationally efficient. Simulations and application to the ABIDE neuroimaging consortium highlight the flexibility of the proposed approach. An R package is provided for ease of implementation.
\end{abstract}

\noindent%
{\it Keywords: Alternating direction method of multipliers, Generalized method of moments, Homogeneity pursuit, Scalable computing.} 

\end{singlespace}

\vfill

\newpage

\section{Introduction}
\label{sec:introduction}

The development of methods to integrate mean regression models is crucial to unlocking the scientific benefits expected from the analysis of massive data collected from multiple sources. The utility of these methods, however, depends on first determining the validity of joint mean regression analysis of multiple data sources \citep{Sutton-Higgins, Liu-Liu-Xie}. Determining mean model parameter homogeneity, which we term mean homogeneity structure, is of fundamental importance to generating meaningful results from data integration. Indeed, substantially erroneous conclusions may ensue from integrating data sources that do not have homogeneous mean structures \citep{Higgins-Thompson}. We propose a new fusion method to learn the mean homogeneity structure of multiple data sources and determine the validity of data integration that delivers two key contributions to the existing literature: (i) the generalization to multivariate generalized linear models from dependent data sources and (ii) a new pairwise fusion penalty that estimates the homogeneity of data sources rather than individual covariates from each data source. 

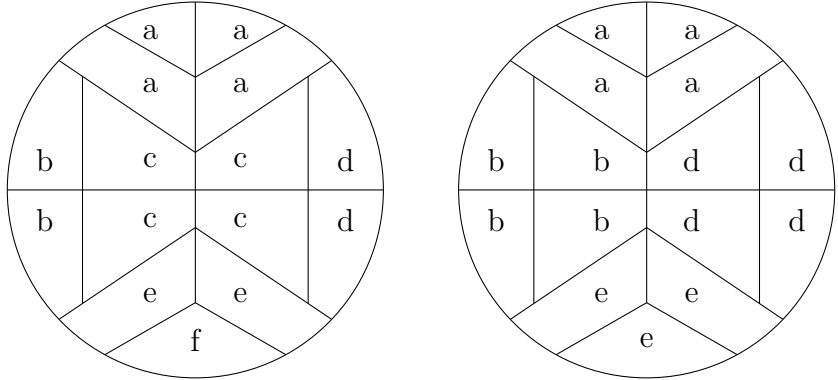
\begin{figure}[h]
\centering
\begin{tikzpicture}
\draw (0,0) circle (2.5cm);
\draw (0,-1.505) -- (0,2.5);
\draw (0,1.5) -- (1.2,2.19);
\draw(0,0.5) -- (1.81,1.72);
\draw(-2.5,0) -- (2.5,0);
\draw(0, -0.5) -- (1.81,-1.72);
\draw(-1.5,1.51) -- (-1.5,-1.51);
\draw(0,-1.5) -- (1.2,-2.19);
\node at (-0.6,2.1) {a};
\node at (-0.6,1.4) {a};
\node at (-2,0.4) {b};
\node at (-0.6,0.4) {c};
\node at (-2,-0.4) {b};
\node at (-0.6,-0.4) {c};
\node at (-0.6,-1.4) {e};
\node at (0,-2) {f};
\begin{scope}[yscale=1,xscale=-1]
\draw (0,1.5) -- (1.2,2.19);
\draw(0,0.5) -- (1.81,1.72);
\draw(0, -0.5) -- (1.81,-1.72);
\draw(-1.5,1.51) -- (-1.5,-1.51);
\draw(0,-1.5) -- (1.2,-2.19);
\node at (-0.6,2.1) {a};
\node at (-0.6,1.4) {a};
\node at (-0.6,0.4) {c};
\node at (-2,0.4) {d};
\node at (-0.6,-0.4) {c};
\node at (-2,-0.4) {d};
\node at (-0.6,-1.4) {e};
\end{scope}
\begin{scope}[shift={(6,0)}]
\draw (0,0) circle (2.5cm);
\draw (0,-1.505) -- (0,2.5);
\draw (0,1.5) -- (1.2,2.19);
\draw(0,0.5) -- (1.81,1.72);
\draw(-2.5,0) -- (2.5,0);
\draw(0, -0.5) -- (1.81,-1.72);
\draw(-1.5,1.51) -- (-1.5,-1.51);
\draw(0,-1.5) -- (1.2,-2.19);
\node at (-0.6,2.1) {a};
\node at (-0.6,1.4) {a};
\node at (-2,0.4) {b};
\node at (-0.6,0.4) {b};
\node at (-2,-0.4) {b};
\node at (-0.6,-0.4) {b};
\node at (-0.6,-1.4) {e};
\node at (0,-2) {e};
\begin{scope}[yscale=1,xscale=-1]
\draw (0,1.5) -- (1.2,2.19);
\draw(0,0.5) -- (1.81,1.72);
\draw(0, -0.5) -- (1.81,-1.72);
\draw(-1.5,1.51) -- (-1.5,-1.51);
\draw(0,-1.5) -- (1.2,-2.19);
\node at (-0.6,2.1) {a};
\node at (-0.6,1.4) {a};
\node at (-0.6,0.4) {d};
\node at (-2,0.4) {d};
\node at (-0.6,-0.4) {d};
\node at (-2,-0.4) {d};
\node at (-0.6,-1.4) {e};
\end{scope}
\end{scope}
\end{tikzpicture}
\caption{Example schematic of 15 brain regions for USA (left) and Europe (right) populations. Regions with the same letter have homogeneous mean structure. \label{data-schematic}}
\end{figure}

This paper is motivated by the Autism Brain Imaging Data Exchange (ABIDE), a consortium of imaging sites across the USA and Europe that aggregated and openly shared neuroimaging data in participants with autism spectrum disorder (ASD) and neurotypical controls \citep{DiMartino-etal}. For each participant in the USA and Europe, summary resting state functional Magnetic Resonance Imaging (rfMRI) outcomes are observed in 15 dependent brain regions. For each group of participants $k \in \{1,2\}$ ($k=1$: USA; $k=2$: Europe), and each brain region $j \in \{1, \ldots, 15\}$, denote by $y_{ir,jk}$ the $r$\textsuperscript{th} neuroimaging outcome in brain region $j$ for participant $i$ in group $k$. The marginal regression model $E(Y_{ir,jk})=\bx^\top_{ir,jk} \bbeta_{jk}$ describes the mean-covariate relationship of interest in brain region $j$ and study $k$, with covariates $\bx_{ir,jk}$ including ASD status. The two central analytic goals are to estimate $\bbeta_{jk}$ and to learn similarities and differences in how the covariates relate to different brain regions in different populations through the homogeneity structure of $\{\bbeta_{jk}\}_{j,k=1}^{15,2}$. An example homogeneity structure is illustrated in Figure \ref{data-schematic}. Learning this structure enables practitioners to leverage homogeneity for improved estimation, and informs whether estimating one model on the combined data, or one marginal model for each brain region and cohort, is appropriate.

More generally in this paper, we consider a complex data integration setting in which multiple independent studies each collect multiple dependent vector outcomes. Potential shared population structures, study design and biological function induce unknown mean structure homogeneity between studies and outcome vectors. Most existing data fusion methods are developed for univariate outcomes \citep{Tibshirani-etal, Tang-Song, Shen-Liu-Xie} and linear models \citep{Li-Nan-Zhu, Ma-Huang, Tang-Xue-Qu} with independent data sources \citep{Ke-Fan-Wu, Wang-Wang-Song-2016}. Approaches developed specifically for longitudinal and spatial data assume working independence between outcomes \citep{Li-Yue-Zhang}. These approaches do not allow for nonlinear modeling, and result in loss of efficiency because they do not incorporate dependence within or between data sources. They also fuse scalar elements of the parameter vector $\bbeta_{jk}$, which results in elements of a parameter vector in a single model being estimated from different data, and fails to provide the desired insights into the shared mean structure of different data sources. There are no suitable fusion methods that can fuse entire mean model parameter vectors $\bbeta_{jk}$, handle multivariate nonlinear models or account for dependence between data sources.

Indeed, a key desired outcome of the ABIDE analysis is to determine the validity of jointly analyzing brain regions and populations. In practice, each data source is traditionally believed to have homogeneous mean across its participants and outcomes, and data sources are integrated as whole units, e.g. \cite{Glass, Xie-Singh-Strawderman}. Existing methods, however, induce a homogeneity partition of covariates that results in estimation of separate elements in $\bbeta_{jk}$ from different data sources. This does not give a clear picture of the validity of integrating data sources. A more useful approach would yield a homogeneity partition of data sources rather than of individual covariate effects. To achieve this, we propose a new formulation of a pairwise fusion penalty that fuses mean model parameter vectors in their entirety, a phenomenon we refer to as model fusion. The resulting estimated homogeneity partition of data sources directly informs the validity of data integrative approaches.

To enable estimation in nonlinear models, we propose to estimate data source-specific mean parameters using a quadratic inference function (QIF) \citep{Qu-Lindsay-Li}. To leverage dependence between data sources, we propose to combine data source-specific QIF and the new pairwise fusion penalty to form a penalized generalized method of moments (GMM) objective function \citep{Hansen, Caner, Caner-Zhang} that non-parametrically estimates dependence between data sources for optimal estimation efficiency. This non-trivial extension requires careful theoretical consideration. Finally, we propose an Alternating Direction Method of Multipliers (ADMM) \citep{Boyd-etal} implementation with an integrated meta-estimator of the fused means in the spirit of \cite{Hector-Song-DIQIF} that optimally weights individual data source estimators. This weighted meta-estimator is asymptotically equivalent to the penalized GMM estimator but more computationally efficient. The resulting fusion method is flexible, efficient and computationally appealing, and can be used, for example, to deliver new insights from massive biomedical studies or as a substitute for meta analysis in the presence of heterogeneity.

The rest of the paper is organized as follows. Section \ref{sec:method} establishes the formal problem setup, describes the QIF construction in each data source and formulates the model fusion objective function. Section \ref{sec:asymptotics} discusses large sample properties. Section \ref{sec:implementation} presents the ADMM implementation details and the integrated meta-estimator. Section \ref{sec:simulations} evaluates the proposed methods with simulations. Section \ref{sec:application} presents the ABIDE data analysis. Proofs, implementation details, additional simulations, ABIDE information and an R package are provided in the Supplementary Material.

\section{Joint Integrative Analysis of Multiple Data Sources}
\label{sec:method}

\subsection{Notation and Problem Setup}
Define $\ba^{\otimes2}$ the outer product of a vector $\ba$ with itself, namely $\ba^{\otimes 2}=\ba \ba^\top$.  Let $(x)_+=x$ if $x>0$ and $(x)_+=0$ otherwise. We consider $K$ independent studies with respective sample sizes $\{n_k\}_{k=1}^K$. In each study we observe $J$ dependent $m_j$-dimensional vector outcomes $\by_{i,jk}=(y_{i1,jk}, \ldots, \allowbreak y_{im_j,jk})$ and covariates $\bx_{ir,jk} \in \mathbb{R}^q$, $r=1, \ldots, m_j$, $j=1, \ldots, J$, for each participant $i=1, \ldots, n_k$ in study $k$, $k=1, \ldots, K$. Here, $\bx_{i,jk}=(\bx^\top_{ir,jk})_{r=1}^{m_j}$ is a $q \times m_j$ covariate matrix assumed to be the study- and outcome-specific observations on the same variables across outcomes and studies. Generalization to participant-specific response dimensions $m_{i,j}$ of $\by_{i,jk}$ is straightforward but omitted for clarity. Participants are assumed independent. This results in a collection of $JK$ data sources that are independent across index $k=1, \ldots, K$ but dependent across index $j=1, \ldots, J$. Such a collection arises, for example, when multiple studies collect multiple dependent outcomes on participants, such as high-dimensional longitudinal phenotypes, pathway-networked omics biomarkers or brain imaging measurements, which collectively form one high-dimensional dependent response vector. 

Consider the generalized linear model for the mean response-covariate relationship of interest, $E(Y_{ir,jk})=\mu_{ir,jk}=h(\bx^\top_{ir,jk} \bbeta_{jk})$, $r=1, \ldots, m_j$, where $\bbeta_{jk} \in \mathbb{R}^q$ the parameter vector of interest. Partial homogeneity of the mean structures of different outcomes is common, for example because of shared biological function (e.g. metabolic pathways) \citep{Hector-Song-DIQIF}. Similarly, partial homogeneity of the mean structures of different studies is common, for example because of similar populations, study designs and protocols \citep{Liu-Liu-Xie}. We posit that there is an unknown partition $\mathcal{P}=\{ \mathcal{P}_g\}_{g=1}^G$ of $\{(j,k)\}_{j,k=1}^{J,K}$ such that $\bbeta_{jk} \equiv \btheta_g$ for all $(j,k) \in \mathcal{P}_g$, for some parameter $\btheta=(\btheta_g)_{g=1}^G \in \mathbb{R}^{Gq}$. Let $\bbeta=(\bbeta_{jk})_{j,k=1}^{J,K} \in \mathbb{R}^{JKq}$, and denote by $\btheta_{g0}=(\theta_{r,g0})_{r=1}^q \in \mathbb{R}^q$ the true value of $\btheta_g$. Let $\left| \mathcal{P}_{\max} \right|= \max_{g=1, \ldots, G} \left| \mathcal{P}_g \right|$ and $\left| \mathcal{P}_{\min} \right|= \min_{g=1, \ldots, G} \left| \mathcal{P}_g \right|$. When $\mathcal{P}$ is known, we define $\bZ \in \mathbb{R}^{JKq \times Gq}$ in the Appendix such that $\bbeta = \bZ \btheta$ and $\bbeta_0=\bZ \btheta_0$, with $\btheta_0=(\btheta_{g0})_{g=1}^G \in \mathbb{R}^{Gq}$ denoting the true value of $\btheta$. Letting $\bZ_{jk}$ the $q$ rows of $\bZ$ corresponding to data source $(j,k)$, we can also rewrite $\bbeta_{jk}=\bZ_{jk} \btheta$. Letting $\bZ_{r,jk}$ the $r$\textsuperscript{th} row of $\bZ_{jk}$, finally we have $\beta_{r,jk}=\bZ_{r,jk} \btheta$.

We wish to estimate the partition $\mathcal{P}$ and to estimate $\btheta_0$ based on all $JK$ sources of information. The proposed solution for estimating $\mathcal{P}$ (and, by extension, $\bZ$) must handle two hierarchical levels of dependence: within data sources $(j,k)$ across $r=1, \ldots, m_j$, and between data sources $(j,k)$ across $j=1, \ldots, J$. The proposed solution handles these two levels of dependence differently by estimating data source parameters using QIF and fusing these parameters using a penalized GMM objective function.

\subsection{Data Source Analysis}
\label{subsec:subgroup}

We first describe the estimating function for $\bbeta_{jk}$ in data source $(j,k)$, $j\in \{1, \ldots, J\}$, $k \in \{1, \ldots, K\}$. QIF \citep{Qu-Lindsay-Li} are a state-of-the-art approach that avoids the specification of nuisance parameters related to second-order moments of $\by_{i,jk}$ by modeling the inverse of the working correlation matrix of $\by_{i,jk}$ as a linear expansion of known basis matrices. Let $\bB_{1,jk}, \ldots, \bB_{s_{jk},jk}$ be a sequence of known basis matrices with elements $0$ and $1$. Let
\begin{equation}
\begin{split}
\bPsi_{jk}(\bbeta_{jk})&=
\frac{1}{n_k} \sum \limits_{i=1}^{n_k} \bpsi_{i,jk}(\bbeta_{jk}) \\
&=\frac{1}{n_k} \sum \limits_{i=1}^{n_k} 
\left( \begin{array}{c} 
\dot{\bmu}^{\bbeta~T}_{i,jk} \bD^{-\frac{1}{2}}_{i,jk} \bB_{1,jk} \bD^{-\frac{1}{2}}_{i,jk} (\by_{i,jk}-\bmu_{i,jk})\\
\vdots \\
\dot{\bmu}^{\bbeta~T}_{i,jk} \bD^{-\frac{1}{2}}_{i,jk} \bB_{s_{jk},jk} \bD^{-\frac{1}{2}}_{i,jk} (\by_{i,jk}-\bmu_{i,jk})
\end{array} \right),\label{block-EE-1}
\end{split}
\end{equation}
where $\bD_{i,jk}$ is the diagonal marginal covariance matrix for participant $i$. The sequence $\{ \bB_{t,jk}\}_{t=1}^{s_{jk}}$ accommodates a broad range of data source-specific correlation structures. The two most common structures, the exchangeable and AR(1), can be approximated with $s_{jk}=2$ basis matrices \citep{Qu-Lindsay-Li}. Since the inverse of an AR($d$) correlation matrix is banded, the AR($d$) correlation structure can be approximated with the linear combination of $s_{jk}=d+1$ basis matrices: the first, $\bB_{1,jk}$ is the identity matrix, and the others, $\bB_{r+1,jk}$, have $1$'s on the $r$th diagonal and 0 elsewhere, $r=1, \ldots, s_{jk}$.

The QIF estimator $\arg \min_{\bbeta_{jk}} \bPsi^\top_{jk}\allowbreak (\bbeta_{jk}) \{\sum_{i=1}^{n_k} \bpsi^{\otimes 2}_{i,jk}(\bbeta_{jk}) \}^{-1} \allowbreak \bPsi_{jk}(\bbeta_{jk})$ is consistent and asymptotically normal under mild regularity conditions. When the working correlation structure is correctly specified by the basis matrix expansion, the QIF estimator is semi-parametrically efficient; even when the working correlation structure is misspecified, it remains consistent, and is efficient within a general family of estimators \citep{Qu-Lindsay-Li}. In addition, it has been shown both theoretically and numerically that estimation efficiency of the QIF estimator is higher than the generalized estimating equations estimator \citep{Liang-Zeger, Song-etal}. Thus, the choice of the correlation structure is not essential to the validity of our approach. The QIF also allows our method to be very flexible and widely applicable since it accommodates a broad range of multivariate outcome distributions. Finally, the QIF provides a natural framework for combining estimating functions from dependent data sources through the GMM \citep{Hansen}. 

\subsection{Joint Integration}
\label{subsec:joint}

After the successful construction of local models for each data source, we formulate an integrated objective function that jointly specifies $\bbeta$ over all data sources. Define $M=\sum_{j=1}^J m_j$, $N=\sum_{k=1}^K n_k$ and the participant group indicator $\delta_i(k)=\mathbbm{1}($participant $i$ is in study $k)$ for $i=1, \ldots, N$, $k=1, \ldots, K$. For participant $i \in \{1, \ldots, N\}$, let
\begin{align*}
\bpsi_{i,k} (\bbeta_k) 
&= \left\{
\bpsi^\top_{i,1k}(\bbeta_{1k}), \ldots, \bpsi^\top_{i,Jk}(\bbeta_{Jk})
\right\}^\top, \\
\bpsi_i(\bbeta)
&=\left\{
\delta_i(1) \bpsi^\top_{i,1}(\bbeta_1), \ldots, \delta_i(K) \bpsi^\top_{i,K} (\bbeta_K)
\right\}^\top,
\end{align*}
where clearly only one $\delta_i(k) \bpsi_{i,k}(\bbeta)$ is non-zero for some $k\in \{1, \ldots, K\}$. Then we can define $\bPsi_N(\bbeta)=(1/N) \sum_{i=1}^N \bpsi_i(\bbeta) $, the stacked vector of estimating functions for all data source parameters over all $JK$ data sources. It is easy to show that
\begin{align*}
\bPsi_N(\bbeta)&=\frac{1}{N} \left\{ \sum \limits_{i=1}^{n_k} \bpsi_{i,jk}(\bbeta_{jk}) \right\}_{j,k=1}^{J,K} =\frac{1}{N} \left\{ n_k \bPsi_{jk}(\bbeta_{jk}) \right\}_{j,k=1}^{J,K} \in \mathbb{R}^{\sum_{j,k=1}^{J,K} (qs_{jk})}.
\end{align*}
Due to the QIF approach to defining $\bPsi_{jk}(\bbeta_{jk})$, $\bbeta \in \mathbb{R}^{JKq}$ is over-identified by the $\sum_{j,k=1}^{J,K} qs_{jk}$ estimating functions in $\bPsi_N(\bbeta)$. To overcome this difficulty, we define a GMM \citep{Hansen} objective function with an added fusion penalty for learning $\mathcal{P}$. Define the sample covariance matrix $\bV_N(\bbeta)=(1/N) \sum_{i=1}^N \{ \bpsi_i(\bbeta) \}^{\otimes 2}$ with row- and column-dimension $\sum_{j,k=1}^{J,K} qs_{jk}$. Let $p_{\delta}(\cdot, \lambda)$ the minimax concave penalty (MCP) \citep{Zhang} with tuning parameter $\lambda \geq 0$, where $\delta>1$ controls the concavity of the penalty function:
\begin{align*}
p_{\delta}(t, \lambda)=\lambda \int_0^{|t|} (1-x/(\delta \lambda))_+~dx.
\end{align*}
Then for each $\lambda$, we define the penalized GMM objective function for $\bbeta$, and the integrated estimator of $\bbeta$, as
\begin{align}
Q_N(\bbeta; \lambda)&=
\frac{1}{2} \bPsi_N^\top(\bbeta) \bV^{-1}_N(\bbeta) \bPsi_N(\bbeta) + 
\sum \limits_{\mathcal{H}} p_{\delta} \left( \sum \limits_{r=1}^q \left| \beta_{r,jk}-\beta_{r,j'k'} \right| ; \lambda \right) , \label{def:integrated-objective}\\
\widehat{\bbeta}_{\lambda}&=\arg \min \limits_{\bbeta} Q_N(\bbeta; \lambda), \label{def:integrated-estimator}
\end{align}
respectively, where $\sum_{\mathcal{H}}$ denotes the sum over $\{(j,k),(j',k') \} \in \mathcal{H}$ and $\mathcal{H}$ is the set of unique data source index pairs defined as\\
\scalebox{0.95}{\parbox{\linewidth}{%
\begin{align*}
&\left\{ \{(j,k),(j',k) \}:~j \in \{1, \ldots, J-1\}, ~j' \in \{j+1, \ldots, J\}, ~k \in \{1, \ldots, K\} \right\} \cup \\
&\left\{ \{(j,k),(j',k')\} :~j,j' \in \{1, \ldots, J\}, ~k \in \{1, \ldots, K-1\}, ~k' \in \{k+1, \ldots, K\} \right\}.
\end{align*}
}}\\
Letting $\widehat{\btheta}_{\lambda}=\{\widehat{\btheta}_g \}_{g=1}^{\widehat{G}_{\lambda}}$ be the distinct values of $\widehat{\bbeta}_{\lambda}=[\{ \widehat{\bbeta}_{jk\lambda}\}_{(j,k) \in \widehat{\mathcal{P}}_{g\lambda}} ]_{g=1}^{\widehat{G}_{\lambda}}$ yields the estimated partition $\widehat{\mathcal{P}}_{\lambda}=\{ \widehat{\mathcal{P}}_{g\lambda} \}_{g=1}^{\widehat{G}_{\lambda}} $, a partition of $\{j,k\}_{j,k=1}^{J,K}$, where $\widehat{\mathcal{P}}_{g\lambda}=\{(j,k) : \widehat{\bbeta}_{jk\lambda}=\widehat{\btheta}_g, j\in \{1, \ldots, J\}, k \in \{1, \ldots, K\} \}$.

The sample covariance $\bV_N(\bbeta)$ nonparametrically estimates and leverages dependence between data sources for improved estimation. Inversion of $\bV_N(\bbeta)$ may be numerically unstable due to large $JKq$ or the choice of an exchangeable data source-specific working correlation structure \citep{Hu-Song, Hector-Song-DIQIF}. In this situation, $\bPsi_N(\bbeta)$ can be replaced by its principal components with non-zero eigenvalues as in \cite{Cho-Qu}. The method proposed in this paper remains unchanged with this substitution.

The penalized GMM was studied in \cite{Caner, Caner-Zhang} with Lasso and elastic net penalties. The penalized QIF has mainly been studied for variable selection \citep{Dziak, Cho-Qu}. The particular form of the penalty in \eqref{def:integrated-objective} bears some similarity to \cite{Ma-etal}, who use the $L_2$ norm inside the penalty. Our proposed penalty yields fusion of entire parameter vectors from different data sources to each other, a phenomenon we refer to as mean model fusion. In contrast, individual penalties for scalar differences, as is common in the literature \citep{Ma-Huang, Yang-Yan-Huang}, fuse individual elements in $\bbeta_{jk}$. Fusing individual elements in $\bbeta_{jk}$ yields estimates based on different data in a single model, which can be difficult to justify and interpret. It is more coherent and interpretable for all parameters in a model to be estimated on the same sample and for the same outcome. Our proposed mean model fusion gives insight into the shared mean structure of the data sources that can validate joint analysis and form the basis for the design of targeted future studies.

\section{Asymptotic Properties}
\label{sec:asymptotics}

In this section, we study the asymptotic properties of the proposed integrated estimator $\widehat{\bbeta}_{\lambda}$ in \eqref{def:integrated-estimator}. Denote $n_{\min}=\min \{n_k\}_{k=1}^K$. We allow $K$, $J$ and $q$ to grow with $M$ and $N$, but assume $q < n_{\min}$ and $\sum_{j,k=1}^{J,K} (qs_{jk}) < N$. There are several practical and technical reasons for these assumptions. The first relates to the problem under consideration: integrating data sources with very large number of covariates increases the sources of potential heterogeneity and learning a homogeneous partition $\mathcal{P}$ is therefore less informative; thus data integration is typically conducted with moderately sized $q$. While somewhat mitigated by the continuously updating weight matrix $\bV_N(\bbeta)$, the practical performance of the GMM is known to suffer when $\sum_{j,k=1}^{J,K} (qs_{jk})$ is large relative to the sample size $N$. This is not directly relevant to the study of asymptotic properties, but influences the settings in which these methods are useful and therefore should be taken into consideration. As we will see, these assumptions also allow us to avoid restrictive assumptions on the outcome distribution. Assuming $q < n_{\min}$ ensures identifiability of data source parameters $\bbeta_{jk}$. Assuming $\sum_{j,k=1}^{J,K} (qs_{jk}) < N$ ensures that $\bV_N(\bbeta)$ is symmetric positive definite as $n_{\min} \rightarrow \infty$.
Denote $\bI_d$ the $d\times d$ identity matrix and $\| \cdot \|$ the $L_2$ norm of a vector or matrix. Define the norms
\begin{align*}
\left\| \ba \right\|_{\infty} &= \max \limits_{1 \leq r \leq d} \left| a_r \right| \quad \mbox{for} \quad \ba=(a_1, \ldots, a_d)^\top \in \mathbb{R}^d,\\
\left\| \bA \right\|_{\infty} &= \max \limits_{1 \leq r \leq d_1} \sum \limits_{s=1}^{d_1} \left| A_{rs} \right| \quad \mbox{for} \quad \bA=\left[ A_{rs} \right]_{r,s=1}^{d_1,d_2} \in \mathbb{R}^{d_1\times d_2}.
\end{align*}

Define the $\{\sum_{j,k=1}^{J,K} (qs_{jk})\}\times (KJq)$ dimensional empirical sensitivity matrix $\bS(\bbeta)=-\nabla_{\bbeta} \bPsi_N(\bbeta)$. Define $\bv(\bbeta)=\lim_{N \rightarrow \infty} Var_{\bbeta_0} \{\sqrt{N} \bPsi_N(\bbeta)\}$. Define the population sensitivity matrix in data source $(j,k)$ $\bs_{jk}(\btheta)=-\nabla_{\btheta} E_{\btheta_0} \{ \bPsi_{jk}(\bZ_{jk} \btheta) \}$ and the population sensitivity matrix $\bs(\btheta)=\{ (n_k/N) \bs_{jk}(\btheta) \}_{j,k=1}^{J,K}=-\nabla_{\btheta} E_{\btheta_0}\{ \allowbreak \bPsi_N (\bZ \btheta) \}$ of respective dimensions $\{\sum_{j,k=1}^{J,K} (qs_{jk})\}\times q$ and $\{\sum_{j,k=1}^{J,K} (qs_{jk})\}\times (Gq)$. Note that the dimensions of $\bS(\bbeta)$ and $\bs(\btheta)$ are different since the derivatives are taken with respect to $\bbeta$ and $\btheta$ respectively.

We define the oracle estimator of $\btheta_0$ when $\mathcal{P}$ and $\bZ$ are known as the GMM estimator
\begin{align}
\widehat{\btheta}_{oracle}&=\arg \min \limits_{\btheta} \frac{1}{2} \bPsi^\top_N(\bZ \btheta) \bV^{-1}_N(\bZ \btheta) \bPsi_N(\bZ \btheta),
\label{def:GMM-oracle-estimator}
\end{align}
and let $\widehat{\bbeta}_{oracle}=\bZ \widehat{\btheta}_{oracle}$. We show in Theorem 1 that the oracle estimator is a consistent estimator of $\bbeta_0$.

\begin{theorem}
\label{thm:GMM}
Suppose assumption (A.1) in the Supplementary Material holds, and $Gq=o(N)$. Then, for some constant $0<C<\infty$,
\begin{align*}
&P\left( \left\| \widehat{\bbeta}_{oracle} - \bbeta_0 \right\|_{\infty} \geq C(KJq)^{-1/2} N^{1/2}  \right)\leq  G q N^{-1}.
\end{align*}
\end{theorem}

It follows from Theorem 1 and GMM theory \citep{Newey-McFadden, Donald-Imbens-Newey, Newey} that, for any matrix $\bH^{1/2} \in \mathbb{R}^{d\times Gq}$ such that $\bH=\bH^{1/2} \bH^{1/2~T}$ has finite maximum singular value, $N^{1/2} \bH^{1/2} \bj^{1/2}(\btheta_0) ( \widehat{\btheta}_{oracle}-\btheta_0 ) \stackrel{d}{\rightarrow} \mathcal{N} (\boldsymbol{0}, \bH)$,
where $\bj(\btheta)=\bj^{1/2}(\btheta)\bj^{1/2~T}(\btheta)=\bs^\top(\btheta) \bv^{-1}(\bZ \btheta) \bs(\btheta)$ is the Godambe information matrix. Assumption (A.1) is standard in estimating function theory and ensures the existence of a unique solution to \eqref{def:GMM-oracle-estimator}, controls the shape of the unpenalized GMM objective function minimized in \eqref{def:GMM-oracle-estimator} and its derivative, and controls their norms over the parameter space. We avoid sub-gaussian tail assumptions that are common in the literature by imposing a stronger constraint on the dimension $q$. This allows us, for example, to study multivariate dependent Poisson outcomes. 

Next, we derive the oracle property of our proposed estimator when $\mathcal{P}$ is unknown in the case of partial heterogeneity between data source parameters. 
\begin{theorem}
\label{thm:minimizer}
Suppose assumptions (A.1)-(A.4) in the Supplementary Material hold, $G \geq 2$, $Gq=o(N)$, $KJq=O(\left| \mathcal{P}_{\min} \right|)$, and $\lambda \gg C(KJq)^{-1/2} N^{1/2}$ with $C$ the constant from Theorem 1. Then there exists a local minimizer $\widehat{\bbeta}_{\lambda}$ defined in \eqref{def:integrated-estimator} such that, as $n_{\min} \rightarrow \infty$, $P(\widehat{\bbeta}_{\lambda}=\widehat{\bbeta}_{oracle}) \rightarrow 1$.
\end{theorem}

The proof is inspired by \cite{Ma-Huang}, with additional work due to different conditions and the fact that the oracle GMM estimator in \eqref{def:GMM-oracle-estimator} has no closed form solution. Combined, Theorems 1 and 2 give the distribution of $\widehat{\bbeta}_{\lambda}$ for fixed $\lambda$. Assumption (A.2) gives eigenvalue bounds in a neighbourhood of $\bbeta_0$. Assumption (A.3) gives a condition on the penalty function that is satisfied by the MCP. Assumption (A.4) bounds the minimal distance between estimators from different partition sets. In Theorem 3, we show that the asymptotic result from Theorem 2 also holds in a fully homogeneous setting, i.e. $G=1$ such that $\left| \mathcal{P}_{\min} \right|=\left| \mathcal{P}_{\max} \right|=KJ$, when $\mathcal{P}$ is unknown.

\begin{theorem}
\label{thm:minimizer-2}
Suppose assumptions (A.1)-(A.3) in the Supplementary Material hold, $G=1$, $KJ=o(N)$ and $q=O(KJ)$. Suppose $\lambda \gg C(KJq)^{-1/2}N ^{1/2}$ with $C$ the constant from Theorem 1. Then there exists a local minimizer $\widehat{\bbeta}_{\lambda}$ defined in \eqref{def:integrated-estimator} such that, as $n_{\min} \rightarrow \infty$, $P(\widehat{\bbeta}_{\lambda}=\widehat{\bbeta}_{oracle}) \rightarrow 1$.
\end{theorem}

\section{Implementation}
\label{sec:implementation}

\subsection{The Alternating Direction Method of Multipliers}
\label{subsec:ADMM}

Direct minimization of the objective function in \eqref{def:integrated-objective} can be analytically and computationally challenging because its derivatives involve three-dimensional arrays and the penalty function depends on all $\bbeta$. A popular implementation to minimize the objective function in \eqref{def:integrated-objective} is the Alternating Direction Method of Multipliers (ADMM) \citep{Boyd-etal}, as in \cite{Tang-Xue-Qu} and \cite{Ma-Huang}. We introduce new parameters $\bgamma=\{ \bgamma_{jkj'k'} \in \mathbb{R}^q, \{(j,k),(j'k')\} \in \mathcal{H} \}$, $\gamma_{r,jkj'k'}=\beta_{r,jk}-\beta_{r,j'k'}$, and reparametrize the objective function in \eqref{def:integrated-objective} as
\begin{align*}
Q_N(\bbeta, \bgamma; \delta, \lambda) = \frac{1}{2} &\bPsi_N^\top(\bbeta) \bV^{-1}_N(\bbeta) \bPsi_N(\bbeta) +  \sum \limits_{\mathcal{H}} p_{\delta}\left( \sum \limits_{r=1}^q  \left| \gamma_{r,jkj'k'} \right|; \lambda \right)\\
&\mbox{with }\beta_{r,jk}-\beta_{r,j'k'}-\gamma_{r,jkj'k'}=0,
\end{align*}
and its minimizer in \eqref{def:integrated-estimator} as $\widehat{\bbeta}_{\lambda}= \arg \min_{\bbeta} Q_N(\bbeta, \bgamma; \delta, \lambda)$. We further introduce Lagrangian multipliers $\bt_{jkj'k'}=\{t_{r,jkj'k'}\}_{r=1}^q$ and define
\begin{align*}
Q^*_N(\bbeta, \bgamma, \bt; \delta, \lambda, \rho)&=Q_N(\bbeta, \bgamma; \delta, \lambda) + \sum \limits_{ \mathcal{H}} \bt^\top_{jkj'k'} (\bbeta_{jk}-\bbeta_{j'k'}-\bgamma_{jkj'k'}) +\\
&~~~~~~~~~~~~~~~~~~~~~~~~\frac{\rho}{2} \sum \limits_{\mathcal{H}} \left\| \bbeta_{jk}-\bbeta_{j'k'}-\bgamma_{jkj'k'} \right\|^2,
\end{align*}
where $\bt=\{ \bt_{jkj'k'} \in \mathbb{R}^{q}, \{(j,k),(j'k')\} \in \mathcal{H} \}$ a $(KJ(J-1)/2+(K-1)J^2)q$-dimensional vector, $\rho$ a learning parameter, and $\delta > 1/\rho$ to ensure convexity of the objective function with respect to each $\gamma_{r,jk,j'k'}$. The integrated estimator is the solution to the unconstrained minimization problem $( \widehat{\bbeta}_{\lambda}, \widehat{\bgamma}_{\lambda}, \widehat{\bt}_{\lambda} ) =\arg \min_{\bbeta, \bgamma, \bt} Q^*_N(\bbeta, \bgamma, \bt; \delta, \lambda, \rho)$. For each $\lambda$, the estimated partition is given by the zero values of $\widehat{\bgamma}_{\lambda}$: if $\widehat{\bgamma}_{jkj'k',\lambda}=\boldsymbol{0}$ then data sources $(j,k),(j'k')$ are combined. This gives estimated partition sets $\widehat{\mathcal{P}}_{g\lambda}$, $g=1, \ldots, \widehat{G}_{\lambda}$.

Details on the ADMM implementation are given in the Supplementary Material. The algorithm alternates between estimating local, data source-specific models and synchronizing the estimates across data sources. Data source-specific estimates can be computed in parallel to further reduce computing time. The procedure can be run on a distributed system in which the main computing node performing the synchronization does not have access to individual data sources, thereby protecting the privacy of individual data sources.

\subsection{The Integrated Meta-Estimator}
\label{subsec:comb-est}

To obtain a fused estimator $\widehat{\btheta}_{\lambda}$ of unique values in $\bbeta$, we could average the values of $\widehat{\bbeta}_{\lambda}=( \widehat{\bbeta}_{jk\lambda} )_{j,k=1}^{J,K}$ across $(j,k) \in \widehat{\mathcal{P}}_{g\lambda}$ for each $g=1, \ldots, \widehat{G}_{\lambda}$. This is statistically inefficient, however, because it does not account for the dependence between $\widehat{\bbeta}_{jk\lambda}$. To overcome this difficulty, \cite{Hector-Song-DIQIF} proposed an integrated meta-estimator for estimators from dependent data sources when the true partition $\mathcal{P}$ of $\{(j,k)\}_{j,k=1}^{J,K}$ is known, which we extend here to the setting when $\mathcal{P}$ is unknown. To give the form of this estimator, define the re-ordered versions of $\bpsi_i$ based on the estimated partition, $\widetilde{\bpsi}_i(\widehat{\bbeta}_{\lambda})= \{ \delta_i(k) \bpsi_{i,jk}(\widehat{\bbeta}_{jk\lambda}) \}_{(j,k) \in \widehat{\mathcal{P}}_{g\lambda},g=1}^{\widehat{G}_{\lambda}}$, and the corresponding sample variability matrix $\widetilde{\bV}_N=(1/N) \sum_{i=1}^N \{ \widetilde{\bpsi}_i(\widehat{\bbeta}_{\lambda}) \}^{\otimes 2}$ and sensitivity matrices,
\begin{align*}
\widetilde{\bS}_{jk}&=-\{ \nabla_{\bbeta_{jk}} \bPsi_{jk}(\bbeta_{jk}) \} \lvert_{\bbeta_{jk}=\widehat{\bbeta}_{jk\lambda}}, \quad \widetilde{\bS}=\mbox{blockdiag} \left\{ \left( \frac{n_k}{N} \widetilde{\bS}_{jk} \right)_{(j,k) \in \widehat{\mathcal{P}}_{g\lambda}} \right\}_{g=1}^{\widehat{G}_{\lambda}}.
\end{align*}

Define the re-ordered estimator $\widetilde{\bbeta}_{\lambda} = \{( \widehat{\bbeta}_{jk\lambda} )_{(j,k) \in \widehat{\mathcal{P}}_{\lambda}} \}_{g=1}^{\widehat{G}_{\lambda}}$. The integrated meta-estimator is given by
\begin{align}
\widehat{\btheta}_{int,\lambda}&=\left( \widetilde{\bS}^\top \widetilde{\bV}^{-1}_N \widetilde{\bS} \right)^{-1} \widetilde{\bS}^\top \widetilde{\bV}^{-1}_N \left\{ \left( \frac{n_k}{N} \widetilde{\bS}_{jk} \widehat{\bbeta}_{jk\lambda} \right)_{(j,k)\in \widehat{\mathcal{P}}_{g\lambda}} \right\}_{g=1}^{\widehat{G}_{\lambda}}.
\label{def:oracle}
\end{align}

The key insight into the construction of $\widehat{\btheta}_{int,\lambda}$ in \eqref{def:oracle} is that dependence between data sources is incorporated using sample Godambe information matrices for dependence between estimating functions. The form in \eqref{def:oracle} does not require access to individual level data to estimate dependence between outcomes directly. We show in Corollary \ref{thm:oracle} that there exists a $\widehat{\btheta}_{int,\lambda}$ that is asymptotically equivalent to the oracle estimator in \eqref{def:GMM-oracle-estimator}.
\begin{corollary}
\label{thm:oracle}
If $G\geq 2$, suppose the conditions from Theorem 2 hold; if $G=1$, suppose the conditions from Theorem 3 hold. Then as $n_{\min} \rightarrow \infty$, there exists a $\lambda$ such that, as $n_{\min} \rightarrow \infty$, $P(\widehat{\btheta}_{int,\lambda}=\widehat{\btheta}_{oracle})\rightarrow 1$.
\end{corollary}

\cite{Hector-Song-JMLR, Hector-Song-DIQIF} assume the partition $\mathcal{P}$ of the $JK$ data sources is known \textit{a priori}, and \cite{Hector-Song-JMLR} further assume that $\mathcal{P}$ is a fully homogeneous partition, i.e. $\mathcal{P}=\{(j,k)\}_{j,k=1}^{J,K}$. These methods are clearly not applicable when $\mathcal{P}$ is unknown. Combined, Theorems 2 and 3 and Corollary \ref{thm:oracle} address this gap in the literature by learning $\mathcal{P}$ and obtaining integrated estimators of the fused coefficients for the estimated partition. In addition, \cite{Hector-Song-JMLR} assumes $q$ is fixed, and \cite{Hector-Song-DIQIF} assumes $J$, $K$ and $q$ are fixed.

The integrated estimator is more efficient than estimators based on each data source alone; see \cite{Hector-Song-DIQIF} and Section \ref{sec:simulations} for theoretical and numerical evidence. The asymptotic distribution of $\widehat{\btheta}_{int,\lambda}$ follows from the asymptotic distribution of $\widehat{\btheta}_{oracle}$: for any matrix $\bH^{1/2} \in \mathbb{R}^{d\times Gq}$ such that $\bH=\bH^{1/2} \bH^{1/2 ~T}$ has finite maximum singular value, there exists a $\lambda$ such that
\begin{align*}
N^{1/2} \bH^{1/2} \left( \widetilde{\bS}^\top \widetilde{\bV}^{-1}_N \widetilde{\bS} \right)^{1/2} (\widehat{\btheta}_{int,\lambda}-\btheta_0) \stackrel{d}{\rightarrow} \mathcal{N} (\boldsymbol{0}, \bH).
\end{align*}

\section{Simulations}
\label{sec:simulations}

We examine the computational and statistical performance of the integrated meta-estimator $\widehat{\btheta}_{int,\lambda}$ for sequences $\lambda=\{\lambda_l\}_{l=0}^L$ in two sets of simulations for logistic and Poisson regression. An additional set of simulations in the linear regression setting is provided in the Supplementary Material. Starting value of $\bbeta$ for $l=0$ is set to the QIF estimators $\arg \min_{\bbeta_{jk}} \bPsi^\top_{jk}(\bbeta_{jk})\{ \sum_{i=1}^{n_k} \bpsi^{\otimes 2}_{i,jk}(\bbeta_{jk}) \}^{-1}\bPsi_{jk}(\bbeta_{jk})$, $j=1, \ldots, J$, $k=1, \ldots, K$; for $l>1$, we use $\widehat{\bbeta}_{\lambda_{l-1}}$ as the starting values for the ADMM procedure with $\lambda=\lambda_l$. We select $\widehat{\lambda}$ from a sequence of $\lambda$ values by minimizing the GMM-BIC of \cite{Andrews}: $\widehat{\lambda} = \arg \min BIC(\lambda)$, where
\begin{align}
BIC(\lambda) = N\bPsi^\top_N(\widehat{\bbeta}_{\lambda}) \bV^{-1}_N(\widehat{\bbeta}_{\lambda}) \bPsi^\top_N(\widehat{\bbeta}_{\lambda}) -\log(N) \left( \sum_{j,k=1}^{J,K} qs_{jk} - \widehat{G}_{\lambda}q \right).
\label{e:GMM-BIC}
\end{align}
In practice, we use the generalized inverse of $\bV_N(\bbeta)$ to overcome numerical instability induced by over-fusion with large $\lambda$. Unless otherwise specified, covariates $\bx_i$ for participant $i$ consist of an intercept and two independent randomly sampled $M$-dimensional dependent Gaussian variables. We estimate $\bbeta$ using our proposed ADMM procedure with MCP parameter $\delta=3$, yielding $\widehat{\bbeta}_{\widehat{\lambda}}$, and compute the resulting combined estimator $\widehat{\btheta}_{int,\widehat{\lambda}}$.

In the first set of simulations, we consider the marginal logistic regression model $\log\{ \mu_{ir,jk}/\allowbreak (1-\mu_{ir,jk}) \}=\bx^\top_{ir,jk} \bbeta_{jk}$, where $\by_i$ is sampled from an $M$-variate dependent Bernoulli distribution using the \verb|SimCorMultRes| R package with data source-specific AR(1) correlation structures. We illustrate the finite sample performance of $\widehat{\btheta}_{int, \lambda}$ in two settings with sample size $N=5000$, $K=2$ and $n_1=n_2=2500$. In Setting I, response dimension is $M=500$ with $J=10$ and $42 \leq m_j \leq 59$ for $j=1, \ldots, J$, the true partition is $\mathcal{P}=\{ \mathcal{P}_g \}_{g=1}^5$ with $\mathcal{P}_1=\{ (1,1),(2,1),(1,2)\}$, $\mathcal{P}_2=\{ (3,1),(2,2),(3,2)\}$, $\mathcal{P}_3 =\{ (4,1),(5,1),(4,2),(5,2)\}$, $\mathcal{P}_4=\{(6,1),(7,\allowbreak 1),(8,1),(6,2),(7,2),(8,2) \}$, $\mathcal{P}_5=\{(9,1),\allowbreak (10,1),\allowbreak (9,2),(10,2) \}$, and corresponding true values are $\btheta_{10}=(-4,1,-2)$, $\btheta_{20}=(4,-1,2)$, $\btheta_{30}=(0.8,0.2,0.6)$, $\btheta_{40}=(1,-2,3)$, $\btheta_{50}=(-1,2,-3)$. In Setting II, response dimension is $M=500$ with $J=5$ and $93 \leq m_j \leq 106$ for $j=1, \ldots, J$, the true partition is a fully heterogeneous partition $\mathcal{P}=\{ \mathcal{P}_g \}_{g=1}^{10}$ and true values are $\btheta_{10}=(2, 1.25, -1)$, $\btheta_{20}=(3.5, -4, -3.25)$, $\btheta_{30}=(-2.5, 3.5, 0.5)$, $\beta_{40}=(-3.25, -3.25, 2)$, $\btheta_{50}=(-1.75, -0.25, -4)$, $\btheta_{60}=(-1, 2, 1.25)$, $\btheta_{70}=\allowbreak (-0.25,\allowbreak  2.75, \allowbreak 3.5)$, $\btheta_{80}=(0.5, 0.5, -0.25)$, $\btheta_{90}=(1.25, -1,\allowbreak  -1.75)$, $\btheta_{100}=(-4, -2.5, \allowbreak 2.75)$.

\begin{figure}[h]
\caption{Logistic regression simulation metrics in Settings I (left) and II (right). Setting I: $M=500$, $J=10$, $42 \leq m_j \leq 59$ for $j=1, \ldots, J$, $\mathcal{P}=\{ \mathcal{P}_g \}_{g=1}^5$. Setting II: $M=500$, $J=5$, $93 \leq m_j \leq 106$ for $j=1, \ldots, J$, $\mathcal{P}=\{ \mathcal{P}_g \}_{g=1}^{10}$. \label{simulations:logistic}}
\includegraphics[width=\textwidth]{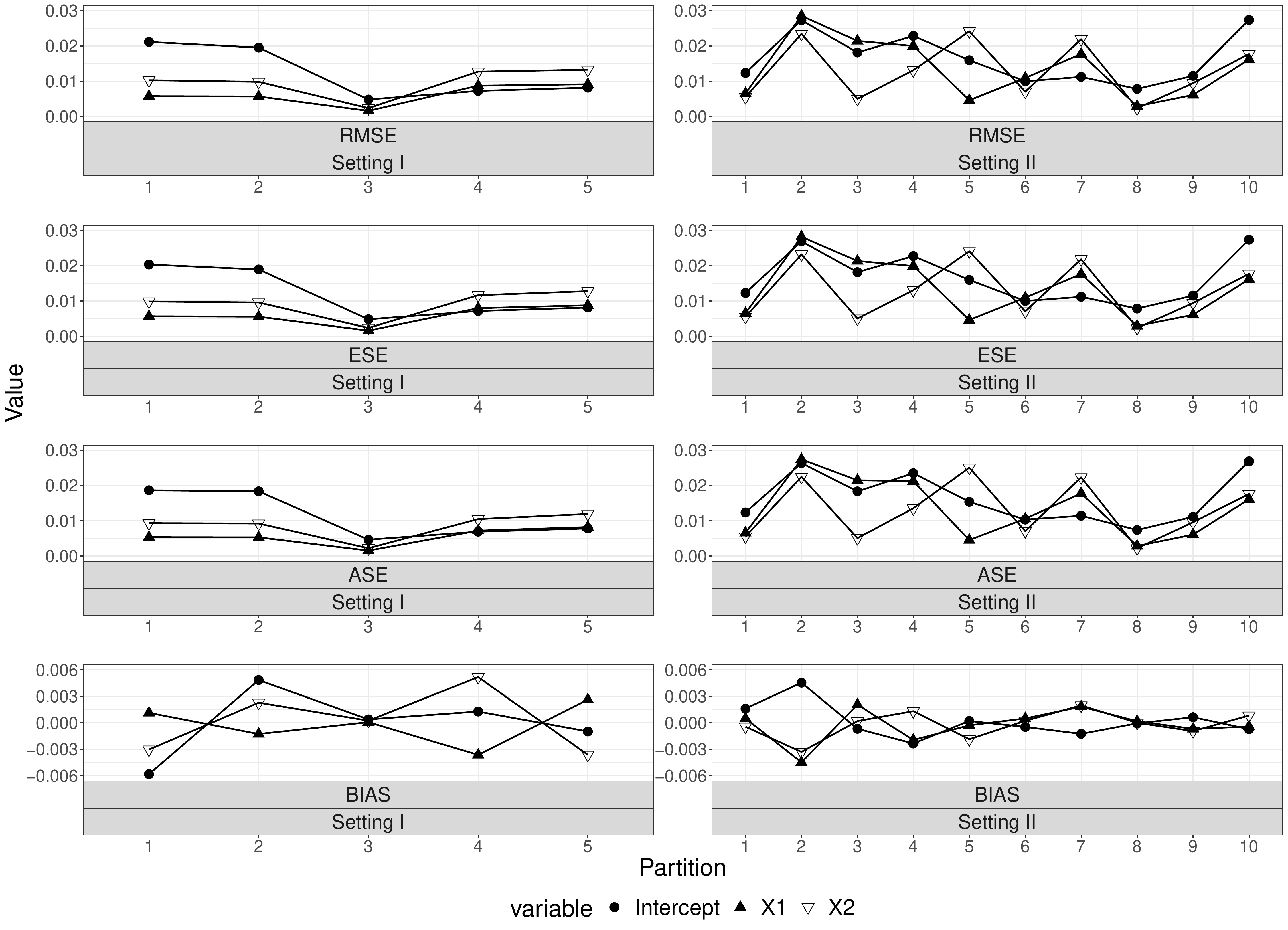}
\end{figure}

We select $\widehat{\lambda}$ from $( 0.05a )_{a=0}^{50}$ using the GMM-BIC criterion in \eqref{e:GMM-BIC}. The selected $\widehat{\lambda}$ recovers the true partition 100\% of the time across 500 simulations in both Settings I and II. Our procedure recovers both partially and fully heterogeneous partitions with high probability. Root mean squared error (RMSE), empirical standard error (ESE), asymptotic standard error (ASE) and mean absolute bias (BIAS) of $\widehat{\btheta}_{int,\widehat{\lambda}}$ averaged over the 500 simulations are visualized in Figure \ref{simulations:logistic}. Consistency of the estimator is illustrated by the near equality of RMSE, ASE and ESE, and the minimal BIAS. Efficiency is improved when a partition set $\mathcal{P}_g$ includes all data sources with common $j$ and $k$ indicators. This is illustrated in Figure \ref{simulations:logistic} by the smaller ASE for sets $\mathcal{P}_3$, $\mathcal{P}_4$ and $\mathcal{P}_5$ in Setting I. This is intuitively justified by the observation that there are more independent samples to estimate these parameters when data sources with common indicators are in the same partition set.

To highlight the benefit of learning the partition $\mathcal{P}$, we compare our approach to a heterogeneous estimator of $\bbeta$. The heterogeneous estimator is the unpenalized GMM estimator $\widehat{\bbeta}_{het}=\bPsi^\top_N(\bbeta) \bV^{-1}_N(\bbeta) \bPsi_N(\bbeta)$, which yields an estimated regression parameter $\bbeta_{jk}$ for each data source $(j,k)$, $j=1, \ldots, J$, $k=1, \ldots, K$. For the intercept, first and second covariate parameters, we average the RMSE, ESE, ASE and BIAS across 500 simulations, and compare these simulation metrics in Table \ref{simulations:comparison} to the similarly averaged RMSE, ESE and BIAS for the intercept, first and second covariate parameter estimates obtained using our fused estimator. In the partially homogeneous Setting I, for the intercept, first and second covariate parameters respectively, RMSE is $1.7$, $2$ and $2$ times smaller for our approach than the heterogeneous estimator, and absolute BIAS is $3.3$, $3$ and $4$ times smaller for our approach than the heterogeneous estimator. In the heterogeneous Setting II, estimating the heterogeneous partition does not result in a loss of efficiency or increase in bias over the heterogeneous estimator, which assumes that the heterogeneous partition is known. Beyond the gain in understanding on the validity of data integration from learning the partition $\mathcal{P}$, we gain significant statistical accuracy and efficiency from learning $\mathcal{P}$ when it is partially homogeneous, and do not lose any desirable properties when it is fully heterogeneous.

In the second set of simulations, we consider the Poisson regression model $\log \mu_{ir,jk}=\bx^\top_{ir,jk} \btheta$. The $M$-variate dependent Poisson response $\by_i$ is sampled by first sampling an $M$-variate Gaussian random variable as in the linear regression simulations, then applying the univariate standard Gaussian cumulative distribution function, and finally applying the Poisson quantile function. We illustrate the finite sample performance of $\widehat{\btheta}_{int,\lambda}$ in two settings with sample size $N=10000$. In Setting I, response dimension is $M=500$ with $J=10$ and $38 \leq m_j \leq 66$ for $j=1, \ldots, J$, and $K=2$ with $n_1=n_2=5000$; the true partition is $\mathcal{P}=\{ \mathcal{P}_g \}_{g=1}^3$ with $\mathcal{P}_1=\{ (1,k),(2,k),(3,k),(4,k) \}_{k=1}^K$, $\mathcal{P}_2=\{ (5,k),(6,k),(7,k) \}_{k=1}^K$, $\mathcal{P}_3=\{ (8,k),(9,k),(10,k) \}_{k=1}^K$ and corresponding true values are $\btheta_{10}=( -0.4,0.1,-0.2 )$, $\btheta_{20}=( 0.1,-0.3,-0.6 )$, $\btheta_{30}=( -0.8,0.2,0.4 )$. In Setting II, response dimension is $M=1000$ with $J=25$ and $27 \leq m_j \leq 53$ for $j=1, \ldots, J$, and $K=1$; the true partition is a fully homogeneous partition $\mathcal{P}= \mathcal{P}_1=\{ (j,k)\}_{j,k=1}^{J,K}$ and true values are $\btheta_0=\btheta_{10}=(0.1, -0.3, -0.6)$.

In both Settings, selecting $\widehat{\lambda}$ in $\{ (0.025a)_{a=0}^{20}, (0.05a)_{a=12}^{40} \}$ as the minimizer of the GMM-BIC recovers the true partition 100\% of the time across 500 simulations. Our procedure recovers both partially and fully homogeneous partitions with high probability. RMSE, ESE, ASE, BIAS and 95\% confidence interval coverage (CP) for $\widehat{\btheta}_{int,\widehat{\lambda}}$ averaged over all simulations in each Setting are reported in Table \ref{poisson-metrics}. Simulation metrics support the inferential properties of the estimator: BIAS is negligible, RMSE, ASE and ESE are approximately equal, and CP\ achieves its nominal level.

\begin{table}[h]
\centering
\caption{Poisson regression simulation metrics in Settings I and II. Setting I: $M=500$, $J=10$, $38 \leq m_j \leq 66$ for $j=1, \ldots, J$, $K=2$, $\mathcal{P}=\{ \mathcal{P}_g \}_{g=1}^3$. Setting II: $M=1000$, $J=25$, $27 \leq m_j \leq 53$ for $j=1, \ldots, J$, $K=1$, $\mathcal{P}= \mathcal{P}_1=\{ (j,k)\}_{j,k=1}^{J,K}$.\label{poisson-metrics}}
\ra{0.9}
\begin{tabular}{llrrrrrr}
Setting & $\mathcal{P}_g$ & Covariate & RMSE$\times 10^{-3}$ & ESE$\times 10^{-3}$ & ASE$\times 10^{-3}$ & BIAS$\times 10^{-4}$ & CP  \\ 
\multirow{9}{*}{I} & \multirow{3}{*}{$\mathcal{P}_1$} & Intercept & $2.20$ & $2.20$ & $2.10$ & $-0.93$ & $0.94$ \\ 
& & $X_1$ & $0.28$ & $0.28$ & $0.28$ & $-0.03$ & $0.94$ \\ 
& & $X_2$ & $0.35$ & $0.35$ & $0.32$ & $-0.11$ & $0.92$ \\ 
& \multirow{3}{*}{$\mathcal{P}_2$} & Intercept & $1.60$ & $1.60$ & $1.60$ & $-1.40$ & $0.95$ \\ 
& & $X_1$ & $0.21$ & $0.21$ & $0.21$ & $-0.18$ & $0.94$ \\ 
& & $X_2$ & $0.29$ & $0.29$ & $0.29$ & $-0.23$ & $0.95$ \\ 
&\multirow{3}{*}{$\mathcal{P}_3$} & Intercept & $2.60$ & $2.60$ & $2.50$ & $-0.49$ & $0.94$ \\ 
& & $X_1$ & $0.24$ & $0.24$ & $0.25$ & $0.07$ & $0.95$ \\ 
& & $X_2$ & $0.35$ & $0.35$ & $0.34$ & $0.19$ & $0.93$ \\ 
\multirow{3}{*}{II} & \multirow{3}{*}{$\mathcal{P}_1$} & Intercept & $0.59$ & $0.58$ & $0.59$ & $-1.00$ & $0.94$ \\ 
& & $X_1$ & $0.08$ & $0.08$ & $0.08$ & $-0.09$ & $0.94$ \\ 
& & $X_2$ & $0.11$ & $0.11$ & $0.11$ & $-0.22$ & $0.94$ \\ 
\end{tabular}
\end{table}

To highlight the benefit of learning the partition $\mathcal{P}$, we compare our approach to the heterogeneous GMM estimator $\widehat{\bbeta}_{het}=\bPsi^\top_N(\bbeta) \bV^{-1}_N(\bbeta) \bPsi_N(\bbeta)$ of $\bbeta$. For each of the intercept, first and second covariate parameters, we average the RMSE, ESE and BIAS across 500 simulations, and compare these simulation metrics in Table \ref{simulations:comparison} to the similarly averaged RMSE, ESE and BIAS for the intercept, first and second covariate parameter estimates obtained using our fused estimator. In the partially homogeneous Setting I, for the intercept, first and second covariate parameters respectively, RMSE is $2.8$, $2.7$ and $2.6$ times smaller for our approach than the heterogeneous estimator. In the fully homogeneous Setting II, for the intercept, first and second covariate parameters respectively, RMSE is $5.8$, $5.4$ and $5.5$ times smaller for our approach than the heterogeneous estimator. We again see the substantial statistical advantage of learning $\mathcal{P}$, especially when the partition is homogeneous.

\begin{table}[h]
\centering
\caption{Comparison between heterogeneous estimator and our approach in the first and second sets of simulations: simulation metric for heterogeneous estimator/simulation metric for our estimator (ratio of metrics). \label{simulations:comparison}}
\subfloat[Logistic model.]{
\ra{0.9}
\begin{tabular}{llrrr}
Setting & Covariate & RMSE$\times 10^{-2}$ & ESE$\times 10^{-2}$ & BIAS$\times 10^{-4}$  \\ 
\multirow{3}{*}{I} & Intercept & $2.07$/$1.22$ ($1.7$) & $2.06$/$1.19$ ($1.7$) & $1.7$/$-0.508$ ($-3.3$) \\ 
& $X_1$ & $1.26$/$0.619$ ($2$) & $1.24$/$0.59$ ($2.1$) & $-6.22$/$-2.1$ ($3$) \\ 
& $X_2$ & $1.91$/$0.97$ ($2$) & $1.89$/$0.924$ ($2$) & $8.66$/$2.17$ ($4$) \\ 
\multirow{3}{*}{II} & Intercept & $1.65$/$1.65$ ($1$) & $1.64$/$1.64$ ($1$) & $0.575$/$1.57$ ($0.4$) \\ 
& $X_1$ & $1.36$/$1.35$ ($1$) & $1.35$/$1.34$ ($1$) & $-3.22$/$-2.69$ ($1.2$) \\ 
& $X_2$ & $1.31$/$1.3$ ($1$) & $1.3$/$1.29$ ($1$) & $-1.88$/$-1.93$ ($1$) \\
\end{tabular}
}\\ \vspace*{1em}
\subfloat[Poisson model.]{
\ra{0.9}
\begin{tabular}{llrrr}
Setting & Covariate & RMSE$\times 10^{-3}$ & ESE$\times 10^{-3}$ & BIAS$\times 10^{-5}$  \\ 
\multirow{3}{*}{I} & Intercept & $5.89$/$2.13$ ($2.8$) & $5.89$/$2.13$ ($2.8$) & $-6.99$/$-9.3$ ($0.8$) \\ 
& $X_1$ & $0.666$/$0.243$ ($2.7$) & $0.665$/$0.243$ ($2.7$) & $-0.724$/$-0.45$ ($1.6$) \\ 
& $X_2$ & $0.859$/$0.332$ ($2.6$) & $0.859$/$0.332$ ($2.6$) & $-0.625$/$-0.514$ ($1.2$) \\ 
\multirow{3}{*}{II} & Intercept & $3.41$/$0.587$ ($5.8$) & $3.41$/$0.579$ ($5.9$) & $-9.68$/$-10.2$ ($0.9$) \\ 
& $X_1$ & $0.419$/$0.0782$ ($5.4$) & $0.419$/$0.0778$ ($5.4$) & $-0.79$/$-0.91$ ($0.9$) \\ 
& $X_2$ & $0.596$/$0.109$ ($5.5$) & $0.596$/$0.107$ ($5.6$) & $-1.92$/$-2.15$ ($0.9$) \\
\end{tabular}
}
\end{table}

\section{Neuroimaging Application}
\label{sec:application}

We illustrate the application of our proposed fusion and estimation procedure with the analysis of the ABIDE I Preprocessed repository rfMRI outcomes \citep{ABIDE} introduced in Section \ref{sec:introduction}. rfMRI measures blood oxygenation in the absence of an external stimulus or task and characterizes intrinsic brain activity \citep{Fox-Raichle}. To study rfMRI outcomes, current standard practice is to fit univariate models to each location, called a voxel, and adjust for multiple comparisons, or to use independent components analysis that treat groups of voxels as independent \citep{Zang-etal, Smith-etal}. These approaches incorporate, at best, partial dependence between outcomes, resulting in loss of power and inability to detect structures of interest in the outcome.

Details on pre-processing, data access and descriptive summary statistics are given in the Supplementary Material. We specify local models in $K=2$ groups of male participants, where $k=1$ is the USA group and $k=2$ the Europe group, and $J=15$ brain regions of interest given in detail in the Appendix. The $J=15$ brain regions are chosen from among the regions of interest defined by the Harvard-Oxford atlas distributed by the FMRIB software library according to published literature on brain regions with amplitude of low frequency fluctuations (ALFF) \citep{Zang-etal} significantly associated with ASD status: left lateral occipital cortex, left temporal occipital fusiform, left occipital fusiform gyrus, right precuneous cortex, frontal pole, superior frontal gyrus, frontal medial cortex, and the superior temporal gyrus \citep{DiMartino-etal, Guo-etal, Li-etal, Mash-etal}. Group sample sizes are $n_1=556, n_2=136$ and response dimensions in brain regions are $(m_1, \ldots, m_J) = (149, 3035, 887, 2105, 372, 1130, 95, 388, 263, 230, 3605, \allowbreak 1187, 1149, \allowbreak 115, 393)$, for a total sample size of $N=692$ and combined response dimension of $M=15103$. For each participant $i$ in group $k \in \{1,2\}$ we consider the mean model $E(Y_{ir,jk})=\beta_{1,jk}ASD_i + \beta_{2,jk} age_i + \beta_{3,jk} IQ_i$, where $Y_{ir,jk}$ is the in-participant centered ALFF response at location $r$ in region $j$ for participant $i$ in group $k$, $ASD_i$ is the centered ASD status ($-0.52$ for ASD and $0.48$ for neurotypical), $age_i$ is the centered age at scan and $IQ_i$ is the centered full-scale IQ.

\begin{figure}[h]
\caption{Solution path for estimated clusters for each $\lambda$ value, with $\widehat{\lambda}$ estimated using the GMM-BIC. 
\label{fig:solution_path}}
\centering
\includegraphics[width=0.85\textwidth]{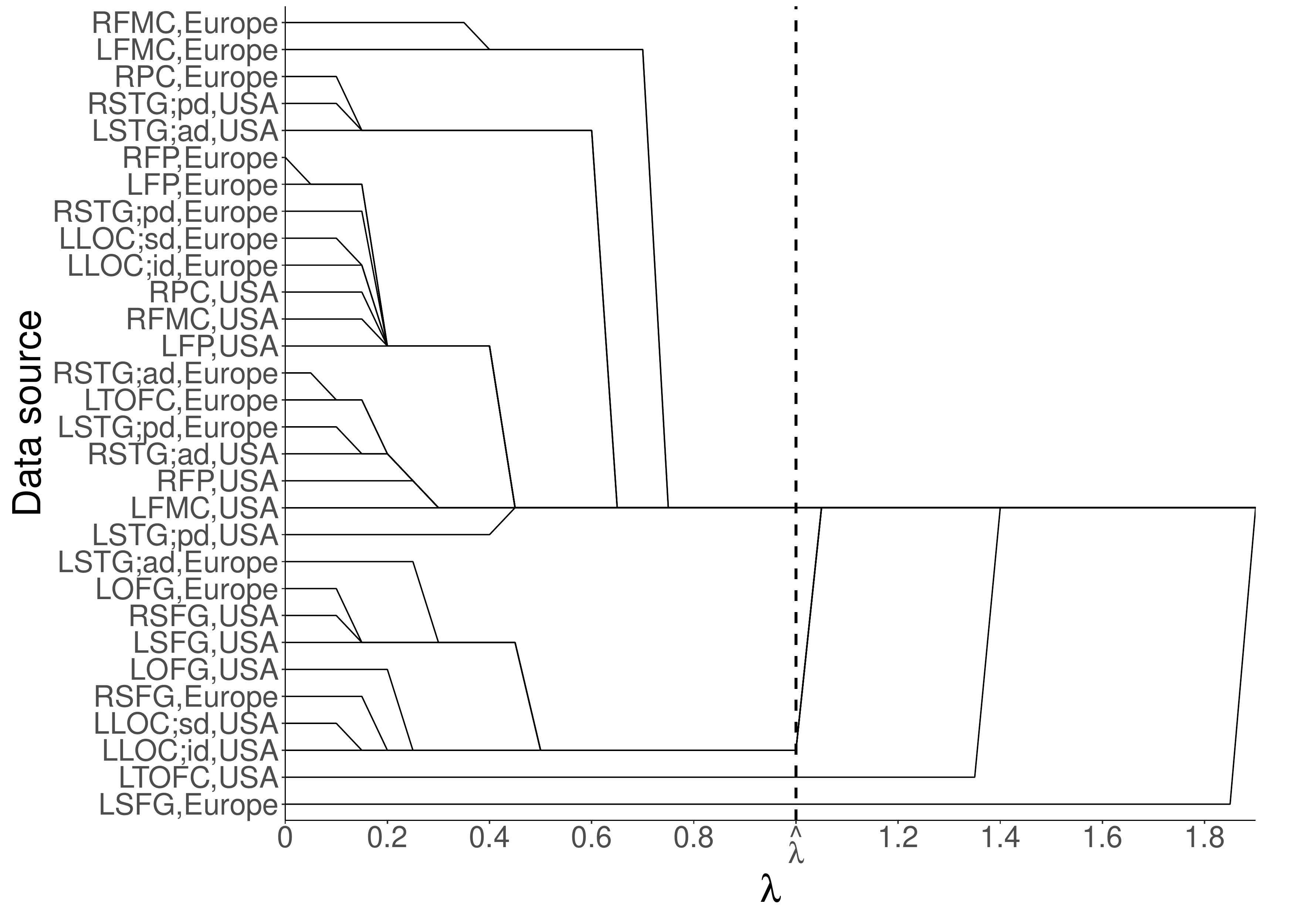}
\end{figure}

Data source mean models with exchangeable working correlation structure are fused over $JK=30$ data sources using the integrated meta-estimator $\widehat{\btheta}_{int,\lambda}$ in \eqref{def:oracle} for $\lambda \in ( 0.05a )_{a=0}^{40}$ with MCP parameter $\delta=3$. Due to numerical instability induced by over-fusion of data sources, we select $\widehat{\lambda}$ as the minimizer of the GMM-BIC in \eqref{e:GMM-BIC} by excluding from consideration $\lambda$ values that yield a homogeneous partition. Minimizing the GMM-BIC gives $\widehat{\lambda}=1$ with $\widehat{G}_{\widehat{\lambda}}=4$. We plot the estimated partition for each value of $\lambda$ and highlight the selected value $\widehat{\lambda}$ in Figure \ref{fig:solution_path}. Unlike the simulations, we do not know the true partition; nonetheless, we have confidence that the GMM-BIC selects an appropriate partition due to the consistent moment selection properties of the GMM-BIC \citep{Andrews} and the simulation results in Section \ref{sec:simulations}. 

Estimated regression coefficients for each partition set are reported in the Supplementary Material. The ASD effect is significant in all partition sets, negative in the first and third partition sets, and positive in the second and fourth partition sets. Two data sources are not fused with any other data source: LTOFC,USA is the only data source in the third partition set (estimate: $-3.8$, 95\% CI: $(-6.0,  -1.6)$), and LSFG,Europe is the only data source in the fourth partition set (estimate: $5.0$, 95\% CI: $(3.6, 6.4)$). One of the most consistent findings in the literature is significantly decreased ALFF in the left middle occipital gyrus for ABIDE ASD versus neurotypical participants \citep{Guo-etal, DiMartino-etal}. Of the Harvard-Oxford regions analyzed, the LLOC corresponds most closely to the left middle occipital gyrus. Interestingly, the LLOC regions are not fused across USA and Europe regions: LLOC,Europe regions exhibit significantly increased ALFF in ASD versus neurotypical participants (estimate: $-1.0$, 95\% CI: $(-1.2, -0.79)$), whereas LLOC,USA regions exhibit significantly decreased ALFF in ASD versus neurotypical participants (estimate: $1.8$, 95\%CI: $(1.4, 2.2)$). It is important to remark here that \cite{Guo-etal} only considered the NYU data collection site, and that the USA participants greatly outnumber the Europe participants, potentially explaining the negative association found by \cite{DiMartino-etal}. Our fusion analysis reveals that the direction of association may be different for USA and Europe populations, which may be due to site, protocol or scanner differences. In addition, the LOFG region overlaps significantly with the left middle occipital gyrus: LOFG,USA and LOFG,Europe are fused with LLOC,USA and exhibit significantly decreased ALFF in ASD versus neurotypical participants. Our fusion analysis thus generally agrees with the existing literature, and is able to tease out subtle differences between data sources. Other regions not fused across USA and Europe cohorts are LSTG;ad, LTOFC and LSFG, indicating additional heterogeneity between USA and Europe in these regions.

\begin{figure}[h]
\caption{Comparison of estimated ASD effects from fused model and individual data sources, with statistical significance (with jitter). \label{fig:ASD_comparison}}
\centering
\includegraphics[width=0.85\textwidth]{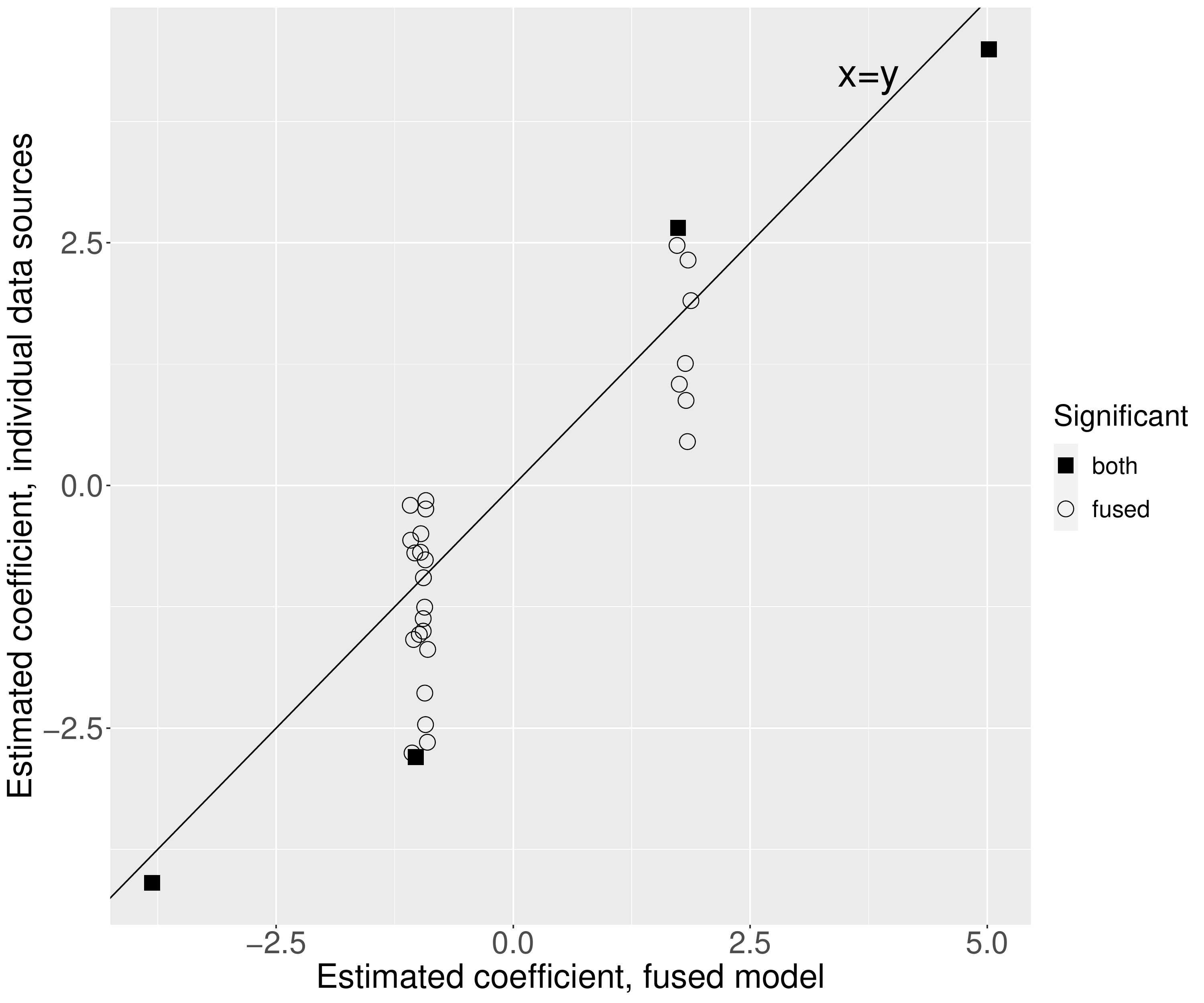}
\end{figure}

We compare estimated ASD effects from the fused model to estimated ASD effects from each individual data source obtained using the heterogeneous GMM estimator $\widehat{\bbeta}_{het}=\bPsi^\top_N(\bbeta) \bV^{-1}_N(\bbeta)\allowbreak \bPsi_N(\bbeta)$ in Figure \ref{fig:ASD_comparison}. As seen in Figure \ref{fig:ASD_comparison}, the estimated fused regression parameter is a weighted average of the individual estimates from each data source; this is the result of the fusion. We observe that the two strongest effects, one positive (LSFG,Europe) and one negative (LTOFC,USA), are not fused with other effects; on the other hand, weaker positive effects are fused together and weaker negative effects are fused together. The fused model results in more efficient estimation of the effect of ASD status, as seen by the increase in number of significant associations. This gain in efficiency is due to the reduction in confidence interval lengths described in Section \ref{sec:simulations}.

At present, there is no consensus on the neurobiology of ASD due to the large body of sometimes contradicting literature; see the review of \cite{Hull-etal}. To improve our understanding of ASD and develop appropriate study designs, it is important to understand the differences and similarities in how ASD impacts different brain regions and populations. We hope that the proposed approach motivates interested readers to explore similar homogeneity structures in the association between ASD and neuroimaging outcomes, for example using other time series summaries than ALFF, or other parcellations of the brain than the Harvard-Oxford atlas.

\section{Discussion}
\label{sec:discussion}

The proposed fused mean structure learning procedure is theoretically sound for application with a broad range of outcome distributions, and simulations illustrate its incredible flexibility and partition recovery.  Fusion of entire mean models from different data sources has not, to our knowledge, been studied before. Most existing literature fuses individual parameters, which results in elements of a parameter vector in a single model being estimated from different data. Our proposed procedure can be modified to perform fusion of individual parameters, but this does not provide insight into the validity of data integration and is more difficult to justify in applications where model interpretation is key. We anticipate the proposed methods will be broadly applicable in the search for ever more powerful integrative analyses of multiple data sources.

Simulations support the inferential properties of the proposed integrated estimator $\widehat{\btheta}_{int,\lambda}$, although practitioners should exercise caution since theoretical results are derived for fixed $\lambda$. Incorporating dependence structure between data sources is primarily to improve statistical efficiency, a concern that is somewhat secondary in selection problems where inference is only valid for fixed $\lambda$. When $\lambda$ is fixed, however, this statistical efficiency is clearly an asset to any estimation procedure. Moreover, the estimating function from each data source over- or just-identifies the parameter $\bbeta$, leading to a natural solution through the GMM formulation that seamlessly incorporates dependence between data sources. Thus, incorporating dependence comes at no cost and can only improve estimation.

Future research directions include fused mean structure learning with multimodal data and allowing $q >n_{\min}$. Multimodal outcomes may follow different distributions so that specifying a valid joint distribution is challenging. Moreover, it is unclear how to define homogeneity coherently across models with different outcome distributions. This interesting topic merits further investigation. Allowing $q>n_{\min}$ is achievable when the regression parameter is sparse. When the sparse subset of covariates is unknown, however, regression parameters are not identifiable from each data source without additional constraints that may introduce bias. \cite{Tang-Zhou-Song-2020} propose to debias data source estimators before integration. Debiasing QIF estimators is an open problem and it is unclear how this debiasing can be incorporated in an iterative procedure that jointly estimates parameters and the homogeneity partition $\mathcal{P}$. See \cite{Yang-Yan-Huang} for related work with continuous univariate outcomes and independent data sources.

\appendix

\section{Appendices}

\subsection{Definition of $Z$}
\label{appendix:Z_def}

When $\mathcal{P}$ is known, let $\bZ^*$ be the $(JK)\times G$ matrix with element $\pi^*_{(j,k)g}$ in column $g$ and row corresponding to data source $(j,k)$ such that $\pi^*_{(j,k)g}=1$ for $(j,k) \in \mathcal{P}_g$ and $\pi^*_{(j,k)g}=0$ otherwise. Define $\bE_{\bbeta} \in \mathbb{R}^{JKq\times JK}$, where the column corresponding to data source $(j,k)$ has 1  in its rows corresponding to the parameters in data source $(j,k)$, and 0 elsewhere. Let $\bE_{\btheta}\in \mathbb{R}^{G\times Gq}$, where the row corresponding to partition set $g$ has 1 in its rows corresponding to the parameters in partition set $g$, and 0 elsewhere. Then let $\bZ=\boldsymbol{E}_{\bbeta} \bZ^* \bE_{\btheta} \in \mathbb{R}^{JKq \times Gq}$. 

\subsection{Brain regions in ABIDE data analysis}
\label{appendix:brain_regions}

We use the following abbreviations for brain regions. Left Frontal Medial Cortex: LFMC. Left Frontal Pole: LFP. Left Lateral Occipital Cortex; inferior division: LLOC;id. Left Lateral Occipital Cortex; superior division: LLOC;sd. Left Occipital Fusiform Gyrus: LOFG. Left Superior Frontal Gyrus: LSFG. Left Superior Temporal Gyrus; anterior division: LSTG;ad. Left Superior Temporal Gyrus; posterior division: LSTG;pd. Left Temporal Occipital Fusiform Cortex: LTOFC. Right Frontal Medial Cortex: RFMC. Right Frontal Pole: RFP. Right Precuneous Cortex: RPC. Right Superior Frontal Gyrus: RSFG. Right Superior Temporal Gyrus; anterior division: RSTG;ad. Right Superior Temporal Gyrus; posterior division: RSTG;pd.

\bibliographystyle{apalike}
\bibliography{fusQIF-bib-20201228}

\end{document}